\documentclass[useAMS,usenatbib]{mn2e}
\usepackage{graphicx}
\usepackage{aas_macros}
\usepackage{amsfonts}
\usepackage{amssymb}
\title[Analytic Central Orbits and their Transformation Group]
  {Analytic Central Orbits and their Transformation Group}
\author[D. Lynden-Bell, S. Jin]
  {D.~Lynden-Bell\thanks{e-mail: dlb@ast.cam.ac.uk} and
    S.~Jin\thanks{e-mail: shoko@ast.cam.ac.uk}\\
  Institute of Astronomy, University of Cambridge, Madingley Road, Cambridge CB3 0HA\\
  Clare College, Trinity Lane, Cambridge CB2 1TL\\
}
 \date{}

\pagerange{\pageref{firstpage}--\pageref{lastpage}} \pubyear{2002}

\def\LaTeX{L\kern-.36em\raise.3ex\hbox{a}\kern-.15em
    T\kern-.1667em\lower.7ex\hbox{E}\kern-.125emX}

\begin{document}

\def\eqnsp{\hspace{-0.25cm}}
\def\deg{\hbox{$\null^\circ$}}

\label{firstpage}

\maketitle

\begin{abstract}
  A useful crude approximation for Abelian functions is developed and
  applied to orbits.  The bound orbits in the power-law potentials $A
  r^{-\alpha}$ take the simple form $(\ell/ r)^k= 1+e \cos(m \phi)$,
  where $k =2-\alpha > 0$ and $\ell$ and $e$ are generalisations of
  the semi-latus-rectum and the eccentricity.  $m$ is given as a
  function of `eccentricity'.  For nearly circular orbits $m$ is
  $\sqrt{k}$, while the above orbit becomes exact at the energy of
  escape where $e$ is one and $m$ is $k$.
  Orbits in the logarithmic potential that gives rise to a constant
  circular velocity are derived via the limit $\alpha\rightarrow
  0$.  For such orbits, $r^2$ vibrates almost harmonically whatever
  the `eccentricity'.
  Unbound orbits in power-law potentials are given in an appendix.
  The transformation of orbits in one potential to give orbits in a
  different potential is used to determine orbits in potentials that
  are positive powers of $r$.  These transformations are extended to
  form a group which associates orbits in sets of six potentials,
  e.g. there are corresponding orbits in the potentials proportional
  to $r, r^{-{2\over 3}}, r^{-3}, r^{-6}, r^{-{4\over 3}}$ and $r^{4}$.
  A degeneracy reduces this to three, which are $r^{-1}, r^2,$ and $
  r^{-4}$ for the Keplerian case.  A generalisation of this group
  includes the isochrone with the Kepler set.
\end{abstract}


\section{Introduction}

Since schooldays when we encountered the rigid pendulum, most of us
have been frustrated by our inability to integrate in elementary terms
Abelian expressions of the form $\int \left[S(u) \right]^{-{1\over2}}
du$, where $S(u)$ has simple zeros at $u_a$ and $u_p \geq u_a$ but is
not quadratic.  In practice $S$ usually depends linearly or
quadratically on parameters which we shall call $\varepsilon$ and $h$,
and its zeros $u_p$ and $u_a$ depend on $\varepsilon$ and $h$ often in
quite complicated ways.  In Appendix A \label{A} we relieve this
frustration by showing how for each pair of $u_a$ and $u_p$, $S$ may be
replaced at lowest order by a quadratic function {\bf{with the same
    zeros}} and the integral may be approximately evaluated
parametrically via perturbation theory.

We do not have to solve $S(\varepsilon, h, u)$ for its zeros $u_p$ and
$u_a$.  Instead we regard $u_p$ and $u_a$ as parameters and then easily
find the $\varepsilon (u_a, u_p)$ and $h (u_a, u_p)$ to which they
correspond.  The process of replacing $S$ by a different quadratic
function for each pair of zeros $u_p$ and $u_a$ we call 
{\bf{quadrating}} (after the old verb `to quadrate' which means `to make
square').  Surely using `quadrate' to mean `to make quadratic' is not
too great an extension!  As the simple, though crude, method developed
can be applied to a far wider class of problems than those encountered
here, we have mentioned it first in the introduction.

Orbits of the general form
\begin{equation}
\label{eq1}
\left( \ell/ r\right)^k = 1+ e \cos \left( m \phi \right)
\end{equation}
have a long history.  Newton in Principia (\citeyear{1687Newton})
showed that orbits of this form with $k=1$ occurred when the central
force was an inverse square law supplemented by an inverse cube force.
His celebrated theorem on revolving orbits demonstrates that if
$r=r(\phi)$ is an orbit of angular momentum $h$ under any central
force $F(r){\hat{\bf{r}}}$, then $r=r(m \phi)$ is an orbit of angular
momentum $m h $ under the central force $\left[ F-\left(m^2-1 \right)
  h^2 r^{-3} \right] {\hat{\bf{r}}}$.  Newton also pointed out that
$r(t)$ was the same for both orbits, so, if the new orbit were viewed
from axes revolving at the rate $(m-1)\dot{\phi}$, then the two orbits
would have the same shape.  But notice that $(m-1) \dot{\phi}$ is not
a constant rotation rate, but speeds up when $r$ is small and slows
down when $r$ is large.  When we view the orbit from axes that rotate
uniformly at the same mean rate $(m-1) <\dot{\phi}>$, the orbits can
have very different shapes involving figures of eight for the more
eccentric ones \citep{1995Notes_RS}.

Recently in a fine paper, \cite{2006AJ....131.1347S}
showed that orbits of moderate or low eccentricity in logarithmic or
power-law potentials with or without cores were well approximated by
analytic orbits of the form (\ref{eq1}).  His approximate orbits are
surprisingly accurate.  Struck was, in part, stimulated to find this
result by a paper by \cite{1997MNRAS.292..905T}
that demonstrated the richness of the resonances in the perturbation
theory of these systems.  \cite{2005MNRAS.358.1273V}
have emphasised that the apsidal precession found in
non-inverse-square orbits is significantly dependent on the
eccentricity of the orbit involved.

Surprisingly, we have been led to orbits of the form (\ref{eq1}) by
looking at orbits of extreme eccentricity $e=1$, where Struck's
methods did not give accurate results.  Standard works on orbits,
\cite{1996thor.book.....B},
\cite{2002ocda.book.....C} and 
\cite{BT87},
do not point out that the orbits of zero energy in power-law
potentials can be exactly solved analytically.  The same variables can
be used to solve the nearly circular orbits.  Since both highly
eccentric and small eccentricity orbits can be so solved, it would be
surprising if there were {\bf{not}} a good approximation, based on the
same variables, that interpolated between $e=1$ and $e=0$.  Struck's
methods do this well for small and moderate eccentricities.  Here we
show that all orbits are well approximated by analytic orbits of the
form (\ref{eq1}) for $0 \leqslant e \leqslant 1$.
 
Here $\ell$ and $e$ are generalisations of the semi-latus-rectum and
the eccentricity.  The potential is $Ar^{-\alpha} = Ar^{k-2}$; this
parameterisation, using $k$ rather than $\alpha$, is chosen in this
section to simplify equations (\ref{eq2}) and (\ref{eq3}) below. If
$r_a$ and $r_p$ are the apocentric and pericentric distances the
generalised eccentricity is given by

\begin{equation}\label{eq2}
e= {\displaystyle{r_p ^{-k} -r_a ^{-k} \over r_p ^{-k}+r_a ^{-k}}= {r_a ^k - r_p ^k
  \over r_a ^k + r_p ^k}}~,
\end{equation}
and the generalised semi-latus-rectum is given by $\ell$ where

\begin{equation}\label{eq3}
\ell^{-k} = {1 \over 2}\left(r_p ^{-k} + r_a ^{-k} \right)~.
\end{equation}

We note that for the Kepler case $k=1$ and the above formulae
all reduce to the usual ones.  We write $\ell = Lr_c$, where $r_c(h)$ is
the radius of the circular orbit of angular momentum $h$.  We have
$r^k _c = (2-k)^{-1} h^2/A$.  The dimensionless parameters $L$ and
$m$ are functions of $e$. Orbits of small eccentricity have
$m=\sqrt{k},$ while those with $e=1$ have $m=k$.  We find the orbits in
the $-V^2 \ln r$ potential from the limiting case $k \rightarrow
2$.

In Appendix~A we show how to improve the accuracy of our orbits via
perturbation theory; however for most purposes the simplicity of the
initial approximation (\ref{eq1}) outweighs the extra complication
that accompanies greater accuracy.  Our methods can be applied to
non-power law potentials (see \cite{1979AJ.....84.1697K}) but here,
for simplicity, we limit ourselves to power laws.

While our methods can be extended to unbound orbits, the results are
less pleasing so they are consigned to Appendix~B.

In section \ref{sec:transf_theory} we use the transformation theory of
Newton, \citeauthor{1911BuAsI..28..113B}, \citeauthor{1990Arnold} and
others to transform our orbits for $0<k<2$ into orbits in potentials
with positive powers of $r$.  We show how that theory can be extended
naturally to give a set of transformations that form a group.  We
develop the subgroup of switch transformations and show that orbits in
the potentials $\psi \propto r^{k-2}$ are conjugate to orbits in the
potentials $r^{2(2-k)/k}, r^{-k}, r^{2k/(2-k)}, r^{-4/k}$ and
$r^{-4/(2-k)}$.  These transformations are not restricted to power
laws, although special simplifications occur for them.  Applications
are made to Plummer's law.

It is shown that the full group has a transformation that connects the
Keplerian potential to the isochrone.

\section{Analytic Orbits}

\subsection{General Orbits in Potentials with ${\bmath{0< k <2}}$}
\label{subsec:gen_pot}

Those looking for orbits in potentials with powers $k$ outside
the above range should consult section \ref{sec:transf_theory}.

A general orbit of specific energy $\varepsilon$ and specific angular
momentum $h$ in the power-law potential $\psi = Ar^{k-2}$ has, in
the usual notation, $r^2 \dot{\phi} = h$ and 
\[
\dot{r}=\sqrt{2
  \varepsilon +2 Ar^{k-2} -h^2 r^{-2}}~.
\]
Now
\begin{eqnarray}\label{eq4}
d \phi = \dot{\phi}dt  =  hr^{-2} dr/\dot{r} = \frac{dr}{r
\sqrt{2 \varepsilon h^{-2} r^2 + 2 Ah^{-2} r^k -1}}\;.
\end{eqnarray}
In place of $r$ we shall use a dimensionless variable which
generalises the $u \left(= {1 \over r}\right)$, so useful in the
Keplerian case:

\begin{equation}\label{eq5}
u=h^2 /(A r^k )\;.
\end{equation}
We also define a dimensionless energy
\begin{equation}\label{eq6}
E=\left( {\varepsilon \over A} \right) \left( {h^2 \over A}
\right)^{(2-k)/k}\;.
\end{equation}
Now $k\; dr/r = -du/u $ so we may rewrite (\ref{eq4}) in terms of $u$:
\begin{equation}\label{eq7}
k~ d\phi = - \left[S(u) \right]^{-{1 \over 2}} du\;,
\end{equation}
where
\begin{equation}\label{eq8}
S(u) = 2Eu^\sigma +2u-u^2\;,
\end{equation}
and $\sigma = 2(k-1)/k$, which is less than one for $0< k < 2$.  
For the marginally bound orbits $\varepsilon =0$, the $u^\sigma$ term
in (\ref{eq8}) disappears so we may integrate (\ref{eq7}) exactly.
Substituting $u=1+ \cos \eta$ reduces (\ref{eq7}) to $k\; d\phi=d \eta$.
Choosing the zero of $\phi$ at that pericentre where $\eta=0$, we have
$\eta=k  \phi$, so the solution for the orbit is of the form of
equation (\ref{eq1}) with $e=1$ and $m=k$:
\begin{eqnarray*}
u=\left( \ell /r \right)^k = 1+ \cos \left( k \phi \right)\;.
\end{eqnarray*}
Indeed, it was this result that motivated our choice of $u$ as the
basic variable.  Nearly circular orbits can also be nicely treated in
terms of $u$, so this encouraged us to conjecture that all bound
orbits can be found analytically to good accuracy.  

Our analytic strategy for integrating equation (\ref{eq7}) more
generally is to replace $S(u)$ with a quadratic function $S_Q(u)$,
which has precisely the same zeros, $u_a$ and $u_p$, corresponding to
the apocentre and pericentre of the orbit.  Thus $S_Q =
q^2(u_p-u)(u-u_a)$ whose $q^2$, the coefficient of $-u^2$ in $S_Q$, is
to be determined so that in some average sense $S_Q(u)$ is a good
approximation to $S(u)$ in the radial range $u_p\ge u \ge u_a$
occupied by the orbit.  For example, we find that choosing $q^2$ so
that $\int_{u_a}^{u_p}S(u)u^{-3/2}du = \int_{u_a}^{u_p}S_Q(u)u^{-3/2}du$
gives a $q$ that is good to $2\%$ accuracy.  A better choice, given
later, is the natural starting point for the perturbation theory of
Appendix~A.

Once $S(u)$ in (\ref{eq7}) has been replaced by the quadratic
$S_Q(u)$, the integration is easy.  From (\ref{eq2}), $e =
(u_p-u_a)/(u_p+u_a)$, so one sets $u_p = \bar{u}(1+e)$ and it
follows that $u_a = \bar{u}(1-e)$ so $S_Q = q^2\left[e^2\bar{u}^2 -
  (u-\bar{u}^2) \right]$.  If we make the substitution $u =
\bar{u}(1+e\cos\eta)$, we find that the integration of (\ref{eq7})
gives $qk\phi=\eta$, so the orbit is
\begin{eqnarray*}
(\ell/r)^k = u/\bar{u} = 1 + e\cos(qk\phi)\;,
\end{eqnarray*}
which is of the form (\ref{eq1}) with $m=qk$.

Crucial to this method of solving for the orbits is the knowledge of
$u_p$ and $u_a$.  A critical step in finding them is to regard the $r_p$
and $r_a$ of an orbit as given,in place of its energy and angular
momentum.  Those can easily be found if $r_p$ and $r_a$ are given, but
solving the other way around is usually difficult.
Once the orbit has been found, this approximation allows us to
determine the radial action and hence the time from pericentre to a
given point on the orbit.  

Having outlined our general procedure, we now turn to solving the
circular and nearly circular orbits using $u$, rather than $r$, as the
variable.  

\subsection{Nearly Circular Orbits}

For the circular orbits of angular momentum $h$, the centrifugal
force balances gravity, so $h^2r_c^{-3}= -d\psi/dr= (2-k)
Ar_c^{k-3}$ and so for them $u=u_c= 2-k$.  Also since
$\dot{r}$ is zero for them their energy is $\varepsilon_c$ where
\begin{eqnarray*}
\varepsilon_c = {\scriptstyle{1 \over 2}} h^2 /r^2_c - Ar^{k-2}_c\;.
\end{eqnarray*}
Also $S(u_c) = 0$, so in our dimensionless variables, {\it{c.f.}}
equation (\ref{eq8}),
\begin{equation}\label{eq9}
E_c = -{\scriptstyle{1 \over 2}} k (2-k)^{(2-k)/k}\;.
\end{equation}

We consider first orbits with energies not much above $E_c$
and we set

\begin{equation}\label{eq10}
\Delta=- \left(E-E_c \right) / E_c\;,
\end{equation}
then $\Delta$ is small for nearly circular orbits and one at the
energy of escape.
\[
E=E_c \left(1-\Delta \right)\;, 
\]
where $E_c$ is given by (\ref{eq9}).
For the nearly circular orbits, we expand the obstreperous $u^\sigma$
term in (\ref{eq8}) about $u=u_c$, omitting terms higher than
quadratic in $u-u_c$:
\begin{eqnarray*}
u^\sigma & = &
u_c^\sigma \left[ 1 +\left({u-u_c \over u_c} \right)
  \right]^\sigma\\ 
 & \simeq & u_c^\sigma \left[1+ \sigma \left({u-
    u_c \over u_c} \right)- {1 \over 2}\sigma (1-\sigma)
  \left({u-u_c \over u_c}\right)^2\right]\;,
\end{eqnarray*}
so inserting this result into (\ref{eq8}),
\begin{eqnarray*}
S(u) \simeq k u_c \Delta +2 \Delta \left(k-1 \right) \left(u-u_c \right) - q^2\left( u-u_c \right)^2\;,
\end{eqnarray*}
where again the coefficient of $-u^2$ in $S_Q$ is $q^2$ and here

\begin{equation}\label{eq11}
q^2= \left[1+ \Delta(k-1) \right]/k\;.
\end{equation}
Completing the square on $(u-\bar{u})$, we have
\begin{equation}\label{eq12}
S \simeq q^2 \left[e^2 \bar{u}^2 - \left(u- \bar{u} \right)^2
  \right]\;,
\end{equation}
where $\bar{u}= u_c + (k-1) \Delta/q^2$ and
\begin{equation}\label{eq13}
e^2 = \Delta q^{-4} \bar{u}^{-2} \left[ u_c k q ^2+ \Delta (k-1)^2
  \right]~.
\end{equation}
Integrating (\ref{eq7}) with $S$ given by (\ref{eq12}) by writing
$u=\bar{u}\left(1+e \cos \eta \right)$
yields 
\begin{equation}\label{eq14}
kq\phi = \eta~;
\end{equation}
so the orbits take the form (\ref{eq1}) with $m=kq$.  Notice that as
$\Delta\rightarrow 0$, $q\rightarrow k^{-1/2}$ and $m\rightarrow\sqrt{k}$.

Whereas these formulae have been derived by neglecting the $(u-
u_c   )^3$ and higher terms in the expansion of $u^\sigma$, it should
be realised that the coefficient of the $u^\sigma$ term itself
vanishes at the energy of escape.  Thus, despite this neglect, our
formulae are exact, not just for $\Delta$ small, but also at
$\Delta=1$.  Indeed at $\Delta=1$ we see that $q^2=1, \bar{u} =1$ and
$e=1$.  However, even such partial reassurance should not deceive us
into believing that formulae (\ref{eq11}) and (\ref{eq13}) are good
enough at intermediate values of $e$.

\subsection{Analysis of General Orbits}
\label{subsec:gen_orbits}

In the non-linear r\'egime, we see from the $e=1$ orbits that $u$
has its mean at $1$ rather than at $2-k$.  It makes little sense to
expand $u^\sigma$ about $u_c=2-k$ when $\Delta$ is not small.
Nevertheless we would like to quadrate $S$, that is, approximate
$S(u)$ by some quadratic function.  We adopt a very different
procedure in the non-linear case.  In place of fixing the energy and
the angular momentum of an orbit and then determining its shape and
size, we choose, instead, a pericentric distance $r_p$ and an apocentric
distance $r_a$.  From these it is simple to find exactly what energy
and angular momentum are needed.  Equivalently we can fix the values
of $\ell$ and $e$ so then $r_p=\ell/(1+e)^{1/k}$ and $ r_a=
\ell/(1-e)^{1/k}$, as can be seen from (\ref{eq1}) with $m \phi$ equal
  to first $0$ and then $\pi$.

Since $\dot{r}=0$ at both $r_p$ and $r_a$, we have for $\varepsilon<0$
\begin{eqnarray}
\label{eq15}
&&\varepsilon + Ar_p^{k-2}-{\scriptstyle{1 \over 2}} h^2 r_p^{-2}=0\;,\\
\label{eq16}
&&\varepsilon + Ar_a^{k-2}-{\scriptstyle{1 \over 2}} h^2 r_a^{-2}=0\;,
\end{eqnarray}
which we may solve for $h^2$ and $\varepsilon$ in terms of $r_a$ and
  $r_p$ or, alternatively, in terms of $\ell$ and $e$:

\begin{eqnarray}\label{eq17}
{h^2 \over 2A} = {r_p^{k-2}-r_a^{k-2} \over r_p^{-2} -
  r_a^{-2}} &=& \ell^k { \left( r_p/ \ell \right)^{k-2}- \left( r_a / \ell
  \right)^{k-2} \over \left( r_p / \ell \right)^{-2} - \left ( r_a / \ell
  \right)^{-2}} \nonumber \\
 &=& \ell^k {\left ( 1+e \right)^{(2-k)/k} - \left( 1-e
  \right)^{(2-k)/k} \over \left(1+e \right)^{2/k}- \left (1-e
  \right)^{2/k}}\;.\nonumber\\
\end{eqnarray}
Multiplying (\ref{eq15}) by $r^2_p$ and subtracting from it $r^2_a
\times$(\ref{eq16}), we deduce
\begin{eqnarray}\label{eq18}
{\displaystyle{-\varepsilon \over A}={r_a^k- r_p^k \over r_a^2-r_p^2}}  
& = & \ell^{k-2} {\left( 1-e\right)^{-1}-\left(1+e \right)^{-1} \over
  \left(1-e \right)^{-2/k} - \left(1+e \right)^{-2/k}} \nonumber\\
& = & \ell^{k-2} {2e \left(1-e^2 \right)^{(2-k)/k} \over
  \left(1+e \right)^{2/k} - \left(1-e \right)^{2/k}}\;.
\end{eqnarray}

\begin{figure}
\includegraphics[width=0.5\textwidth]{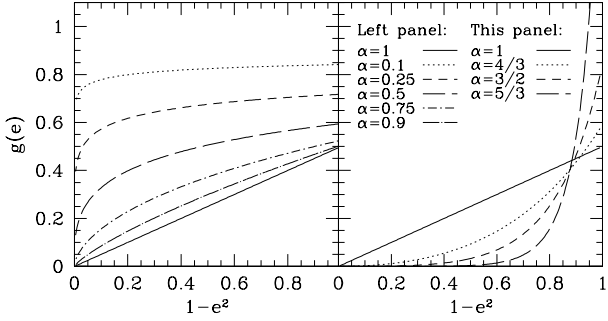}
\caption{The function g(e) for various values of $\alpha = 2-k$.  g(e) is
  minus the dimensionless energy $E$.  As $\alpha \rightarrow 0$, the
  graph tends to the line $g(e) \equiv 1$.}
\label{fig:g_e}
\end{figure}

Another alternative, which is the most useful one in the equivalent
problem in quantum mechanics, is to consider $h$ and $e$ as given.
Then, eliminating $\ell$ in (\ref{eq18}) in favour of $h$ as found from
(\ref{eq17}), we obtain
\begin{eqnarray*}
{-\varepsilon \over A} = \left({h^2 \over A}
\right)^{(2-k)/k}g(e)~;\ E= -g(e)\;,
\end{eqnarray*}
\smallskip
where, setting $\gamma = (2-k)/k$,
\begin{eqnarray}\label{eq19}
 g(e) = 2^{2/k}e(1-e^2)^{\gamma} 
\frac{\left[(1+e)^{\gamma} -(1-e)^{\gamma}\right]^{\gamma}}
{\left[(1+e)^{2/k}-(1-e)^{2/k}
  \right]^{2/k}}\;.
\end{eqnarray}
Despite its strange appearance, $g(e)$ is not a complicated function.
For 
$k=1$
it is ${1 \over 2}(1-e^2)$
and for
$k=2$
it is 1.  We plot $g$ against 
$1-e^2$ 
for several $k$ values in figure~\ref{fig:g_e}.

We wish to approximate $S$ by a quadratic in $u$ which must vanish at
$u=u_p$ and $u=u_a$, so it has to take the form
\begin{equation}\label{eq20}
S\simeq S_Q = q^2 \left(u-u_a \right) \left(u_p-u \right) = q^2 \left[e^2
  \bar{u}^2 - \left(u- \bar{u} \right)^2 \right]\;,
\end{equation}
where $q$ is yet to be determined and has been given that notation to
conform with our earlier definition that $-q^2$ is the coefficient of
$u^2$ in $S$.  In the above,
\begin{eqnarray}\label{eqn:ubar}
\bar{u}= {1 \over 2} \left(u_a + u_p \right) = {h^2 \over 2A}
\left(r_p^{-k} + r_a^{-k} \right) = {h^2 \over A} ~ \ell^{-k}~,
\end{eqnarray}
from which we see that $\bar{u}$ is twice the final expression in
(\ref{eq17}) but without the $\ell^k$.  The `eccentricity' $e$ is $\left(u_p - u_a
\right)/\left(u_p + u_a \right)$ as always.  Using (\ref{eq20}) for
$S(u)$ with the substitution $u=\bar{u} \left(1+e \cos\eta \right)$,
we readily integrate equation (\ref{eq7}) to obtain $qk \phi = \eta$
as before.  So the solution is still equation (\ref{eq1}) with $m=qk$,
but we must still determine $q^2$.  

A useful approximate formula, good to about $2\%$, is given by setting
$\int_{u_a}^{u_p}S(u)u^{-3/2}du = \int_{u_a}^{u_p}S_Q(u)u^{-3/2}du$.  This
gives, for $\sigma\ne 1/2$ (i.e. $\alpha\ne 2/3$),
\begin{eqnarray}\label{eq21}
&&\hspace{-.7cm}
q^2 = \frac{\frac{1}{\bar{u}}\left(2-(\sigma-\frac{1}{2})^{-1}\right)
  + \left((\sigma-\frac{1}{2})^{-1}-\frac{2}{3}\right)
\left(1+\frac{1}{2}\sqrt{1-e^2}\right)}
{\frac{4}{3}\left(1-\sqrt{1-e^2}\right)},\nonumber\\
&&
\end{eqnarray}
with $m=qk$ as before and $\bar{u}$ can be expressed as a function of
$k$ and $e$ only, via (\ref{eq17}) and (\ref{eqn:ubar}).  We have
chosen this power of $u$ in the integrals we equate above, as it gives
the best agreement to the true $m$ without compromising on the
simplicity of the expression for $q$.  We note that choosing $u^{-1}$
in the integrals gives as good an agreement as using $u^{-3/2}$, but
with the resulting $m$ becoming overestimates on the true value for
$\alpha<1$ and vice versa for $\alpha>1$.  Choosing an exponent
between $-1$ and $-1.5$ results in better agreement still, but we lose the
simplicity of the resulting analytic expression for $q$.  The exponent
of $-4/3$ is as good as any, but gives the following somewhat awkward
result:
\begin{eqnarray*}
q^2 = \frac{5(\sigma-1)(e_++e_-)\bar{u}^{-1}+(2-\sigma)\left[2(e_++e_-)+e_+^2e_-^2 \right]}{(3\sigma-1)(e_++e_--2e_+^2e_-^2)}\;,
\end{eqnarray*}
where $e_{\pm} = (1\pm e)^{1/3}$.

The angle between successive apocentres is important as such angles
accumulate as the orbit is prolonged.  We now determine $q$ to get this
angle as accurately as possible.  It is given by 
\begin{eqnarray}
\Phi = \frac{2\pi}{m} = \frac{2}{k}\int^{u_p}_{u_a}S^{-1/2} du
\eqnsp 
&=& \eqnsp
\frac{2}{k}\int^{u_p}_{u_a}\left(\frac{S_Q}{S}\right)^{1/2}\frac{du}{S_Q^{1/2}}
\nonumber\\ 
&=& \eqnsp
\frac{1}{m}\int^{\pi}_{-\pi}\left(\frac{S_Q}{S}\right)^{1/2} d\eta\;
\nonumber,
\end{eqnarray}
hence the average of $(S_Q/S)^{1/2}$ over $\eta$ must be one.  We
evaluate this average over eight points around $-\pi<\eta<\pi$.  There
is a difficulty in evaluating $(S_Q/S)^{1/2}$ exactly at the apocentre
when $e=1$ since the apocentre is at infinity.  Surprisingly, the
result of taking the limit of $r_a$ as $e\rightarrow 1$ gives a
different (and wrong) result from setting $e=1$ and then evaluating
the limit as $\cos\eta\rightarrow -1$.  We get around this by using
$\cos\eta=-0.990$, where everything is finite, in place of $\eta=\pi$.

At pericentre $u_p$, both $S_Q$ and $S$ are zero but the limit of
${\scriptstyle{1\over q}}(S_Q/S)^{1/2}$ is 
\begin{eqnarray*}
\sqrt{ \frac{2\bar{u}e}{\sigma(2-u_p)-2(u_p-1)}}\;.
\end{eqnarray*}
We evaluate 
\begin{eqnarray*}
\frac{1}{q}\sqrt{\frac{S_Q}{S}} = 
\sqrt{\frac{(u_p-u)(u-u_a)}{2Eu^\sigma + 2u - u^2}}\;,
\end{eqnarray*}
where $u=\bar{u}(1+e\cos\eta)$, at the other seven points $\pm\pi/4$,
$\pm\pi/2$, $\pm3\pi/4$ and $\cos\eta=-0.990$.  
For convenience, we label the values of $u$ at $\pm\pi/4$ and $\pm
3\pi/4$ as $u=\bar{u}(1+e/\sqrt{2}) = u_+$ and
$u=\bar{u}(1-e/\sqrt{2})=u_-$ respectively.  
Our estimate of $1/q$
is the average over the eight values that result:
\begin{equation}\label{eq22}
\frac{1}{q} = \frac{1}{8}\displaystyle\sum_{i=1}^{8} \left[\frac{(u_p-u_i)(u_i-u_a)}{S(u_i)}\right]^{1/2}\;;\;\; m=kq\;.
\end{equation}
At each $\alpha$,
the resulting $q$ is a (somewhat complicated) function of $e$, since
$E$, $u_a$, $u_p$, $\bar{u}$ are all functions of $e$.

\subsection{Comparisons with Computed Orbits}

\begin{figure}
\includegraphics[width=0.5\textwidth]{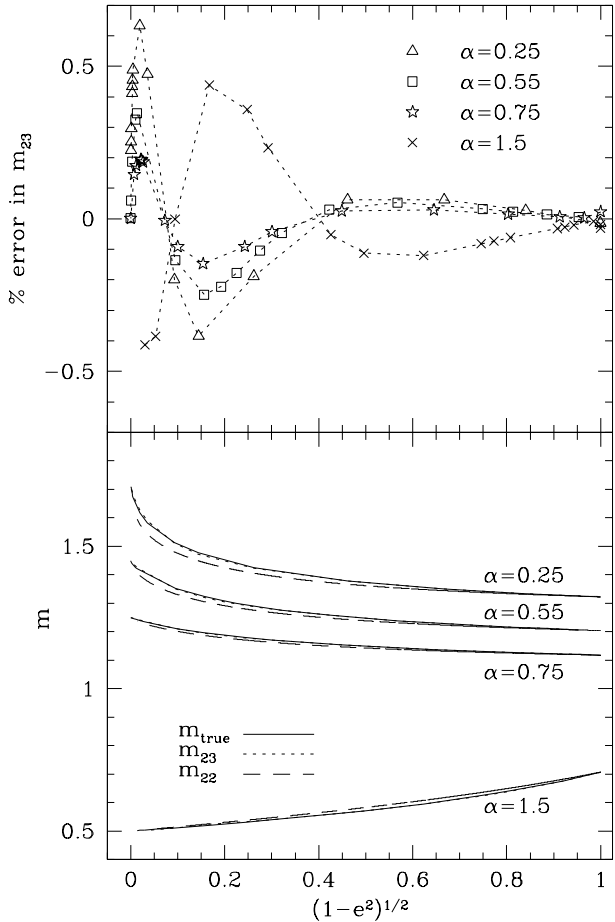}
\caption{The top panel shows the percentage difference between the
  true value of $m$ from computed orbits and $m_{23}$ calculated from
  the analytical formula (\ref{eq22}), plotted as functions of
  $\sqrt{1-e^2}$ for different values of $2-k=\alpha$.  The largest
  deviations occur at high eccentricities; the region 0.1 to 0.3 in
  $\sqrt{1-e^2}$ corresponds to $0.995>e>0.954$.  The bottom panel
  shows the values of $m$, estimated using (\ref{eq21}) and
  (\ref{eq22}), along with the true values, again for four values of
  $\alpha$.  }
\label{fig:mfig}
\end{figure}

\begin{figure}
\includegraphics[width=0.5\textwidth]{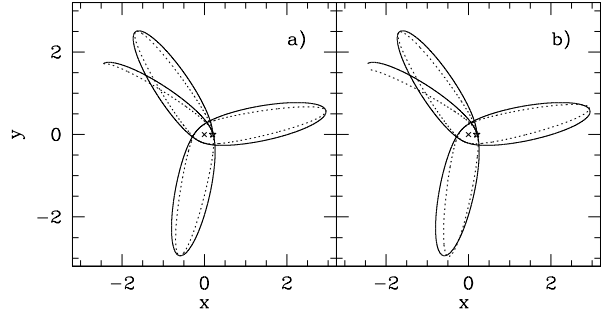}
\caption{An orbit in the $2-k=\alpha=0.25$ potential, chosen to be in the
  high eccentricity region where the approximation is least good.  The
  full line is the computed true orbit.  In panel a), it is compared
  to $(\ell/r)^k = 1+e\cos(m\phi)$ (dotted line) with $m$ chosen to
  agree with the computed orbit.  This shows the deviation in the
  shape of the lobes.  In panel b), the value of $m$ is estimated by
  the 8-point average of equation (\ref{eq22}) and the error in $m$ is
  readily seen from the different precession of the solid and dotted
  orbits.  The difference in the true value of $m$ and that estimated
  from the 8-point average is less than $0.5\%$.  All orbits start at
  the starred location.}
\label{fig:orbcompar}
\end{figure}

\begin{figure}
\includegraphics[width=0.5\textwidth]{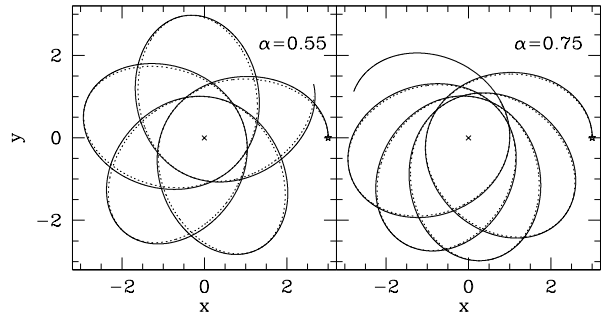}
\caption{Two computed orbits with the same $r_{\rm{min}}/r_{\rm{max}}$
in potentials with different $2-k=\alpha$ are compared with the analytic
orbits $(\ell/r)^k = 1+e\cos(m_{\rm{true}}\phi)$ (dotted lines).  The $\alpha =
0.55$ orbit has $e=0.662$ while the $\alpha=0.75$ orbit has $e=0.596$
and precesses much less rapidly.  Both the shapes and the precession
rates of these orbits are well represented by the analytic formula.}
\label{fig:rosette}
\end{figure}

\begin{figure}
\includegraphics[width=0.5\textwidth]{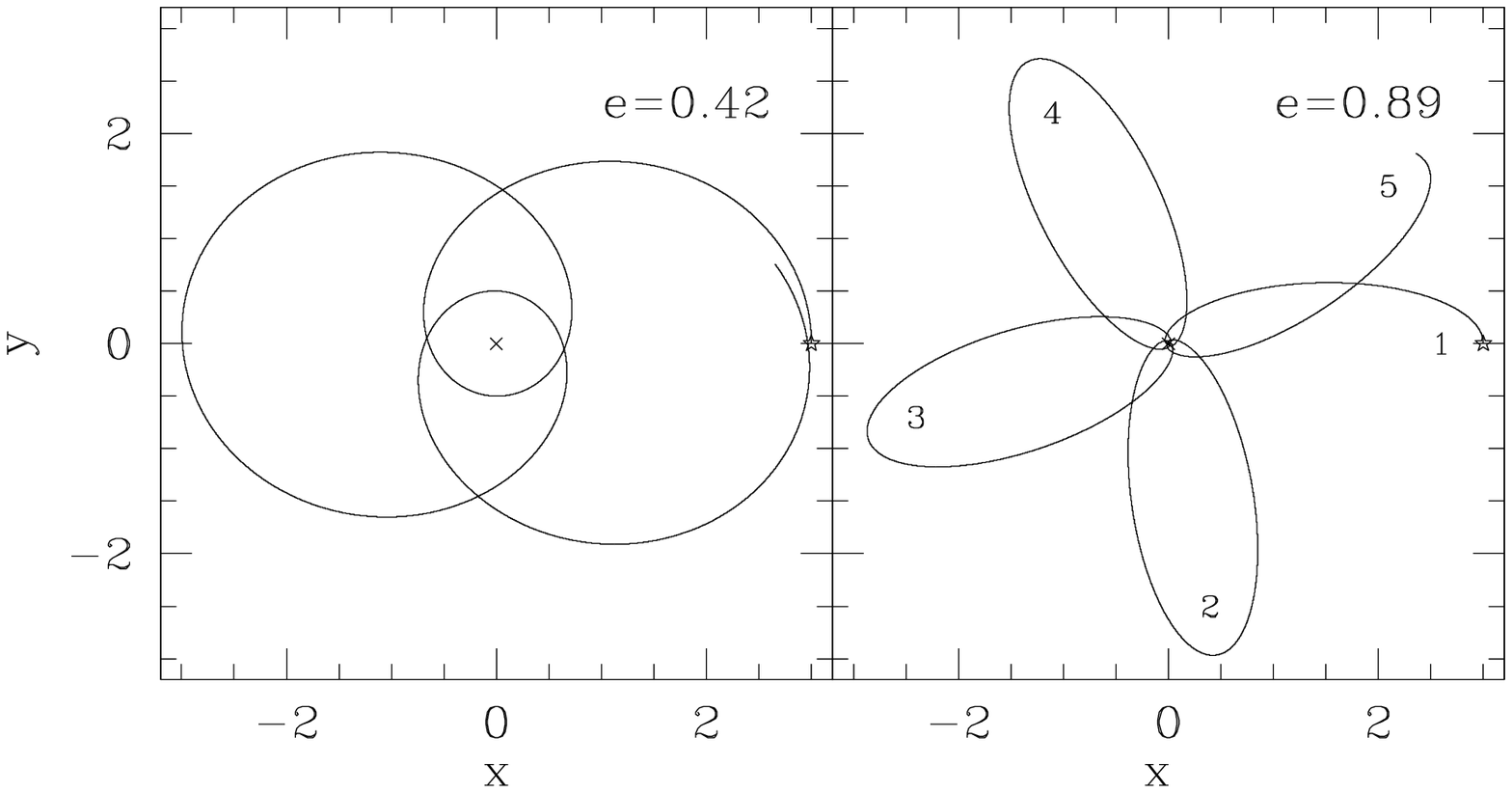}
\caption{Two orbits in the $2-k=\alpha=1.5$ potential. The orbit on the left has
a forward precession by nearly $180\deg$ whilst that on the right has
a forward precession by a little over $270\deg$.  The turn close to
the origin in the latter cannot be seen but is a more rapidly turning
version of that seen on the left.  The lobes are numbered in
sequential order for the higher eccentricity orbit.}
\label{fig:alpha1.5}
\end{figure}

Orbits were computed in the $x, y$ plane from the Cartesian form of
the equations of motion for potentials with $\alpha= 0.25,~ 0.55,~
0.75,~ 1.5 ~ $ and $1.0$.  The last provides a valuable check that we
get $m=1$ in the Newtonian case, even at very high eccentricities of
order 0.999.  We also checked that $m= \sqrt{k}$ for nearly circular
orbits and that $m \rightarrow k$ as $e \rightarrow 1$ for all the
values of $\alpha$.  As $m$ varies quite rapidly with eccentricity as
$e \rightarrow 1$, accurate computations are required at high
eccentricities.  Orbits in the logarithmic potential which gives a
constant circular velocity are considered later, in section
\ref{subsec:log_pot}; the approximation adopted there is somewhat
different.

Figure \ref{fig:mfig} shows a comparison between our estimated values
of $m$ and the computed values of $m$ for potentials with
$\alpha=0.25$, 0.55, 0.75 and 1.5.  For all values of $\alpha$ shown,
the deviation of $m$ calculated from the analytical formula (\ref{eq22})
is only a fraction of a percent.  Notice that both plots are against
$\sqrt{1-e^2}$ so that high eccentricities are on the left and low
eccentricities are on the right.  It is, of course, possible to read
off $m$ as a function of $\sqrt{1-e^2}$ or of $e$ from the computed
points in this figure.

Panel a) of figure \ref{fig:orbcompar} shows a computed orbit in the
potential with $\alpha = 0.25$ together with an orbit of the same $m,~
r_a $ and $ r_p$ but calculated from the equation $\left( \ell/r
\right)^k = 1 + e \cos (m \phi)$.  This demonstrates how the shape
given by equation (\ref{eq1}) fits the computed orbit.  A better fit
is obtained using the perturbation theory of Appendix A.
The dotted orbit in panel b) is $ \left( \ell/r \right)^k= 1+e\cos(m_{\ref{eq22}}
\phi)$ and the gradual precession due to the error in the estimated
$m_{\ref{eq22}}$ is readily seen.

Figure \ref{fig:rosette} shows two orbits with the same ratio of $r_a/r_p$ but
in the potentials with $\alpha =0.55$ and $\alpha = 0.75$.  Because
the definition of `eccentricity' we gave in equation (\ref{eq2})
depends on $\alpha$ (through $k$), these orbits have eccentricities of
0.662 and 0.596 respectively.  Notice that the two drawings have the same
number of apsides but these have precessed much less for the  $\alpha =
0.75$ orbit as the potential is closer to the Keplerian $\alpha =1$.

Figure \ref{fig:alpha1.5} shows two orbits in the $\alpha=1.5$
potential for which the precession is forwards because $\alpha>1$
whereas the other illustrations all have a backward precession.  Under
the transformation $\zeta =z^{k/2}$ considered in section
\ref{sec:transf_theory}, these orbits transform into ones in the
potentials $\psi \propto r^{2 \alpha/k} = r^6$.  For orbits with
$\alpha>1$, the straight perturbation theory giving $m=kq$ with $q$
given by (\ref{eq22}) yields $m$ to better than $0.5\%$.

It is often useful to have a vectorial way of delineating orbits and
the velocities of particles describing them.  To do this, we generalise
Hamilton's eccentricity vector which has magnitude $e$ and points toward
pericentre.  As our pericentres precess within the orbital plane, we
invent a rotating eccentricity vector.  If we start at pericentre with
${\bf{e}}={\bf{e}}_0 $ we take ${\bf{e}}$ at later
times to be given by ${\bf{e}}={\bf{e}}_0 \cos \left[\left(1-m \right) \phi
  \right]+ {\bf{\hat{h}}} \times
{\bf{e}}_0\sin\left[\left(1-m\right) \phi \right]$. 
This ${\bf{e}}$ obeys
$d {\bf{e}} / d \phi = \left( 1-m \right) {\bf{\hat{h}}}
\times {\bf{e}}$.
The angle between the radius vector to the particle and the
eccentricity vector is then 
$m \phi$
and the equation of the orbit (\ref{eq1}) can be rewritten 
\begin{equation}\label{eq23}
\left(\ell/r \right)^k = 1+{\bf{e}}.{\bf{\hat{r}}}~.
\end{equation}
The transverse velocity of the particle is clearly 
${\bf{h}} \times \hat{\bf{r}}/r$
and the radial velocity can be obtained from the orbit and the energy
equation.  Using our approximations quadrating the latter, we find
\begin{equation}\label{eq24}
{\bf{v}}= r^{-1} \left[ {\bf{h}} \times {\bf{\hat{r}}}+
  \ q\left(r/\ell\right)^k {\bf{h}}.\left({\bf{e}}\times
  {\bf{\hat{r}}}\right){\bf{\hat{r}}} \right]~.
\end{equation}

Given $\bf{v}$ and $\bf {r}$ at one time one might wish
to use these equations at a later time.  Then one needs to find
$\bf{e}$, $\bf{h}$, $\ell$, $q$ and $m$
from the initial $\bf{v}$ and $\bf{r}$
together with the known potential
$\psi =Ar^{-\alpha}$.
From $\bf{v}$ and $\bf{r}$
it is easy to construct
$\varepsilon={1 \over 2} v^2-Ar^{-\alpha}$ and
$\bf{h}=\bf{r}\times \bf{v}$,
from these $E$ is found.  For given $E$ and $\alpha, e$ may be found
from figure \ref{fig:g_e}.  $\ell$ then follows from (\ref{eq17}) and $m,q $
from (\ref{eq22}).  The direction of $\bf{e}$ within the plane
perpendicular to $\bf{h}$ then follows from
${\bf{e}}.{\bf{\hat{r}}} = \left( \ell/r \right)^k-1$ with
the ambiguity in angle resolved from
${\bf{v}}.{\bf{\hat{r}}}= r^{-1} q \left( r/\ell \right)^k
{\bf{h}}. \left( {\bf{e}} \times {\bf{\hat{r}}}\right)$
which follows from the $\bf{v}$ equation above.  Thus all the
orbital parameters are determined.

\subsection{Action, Adiabatic Invariants and Time}
So far we have concentrated on the shape of the orbit in space,
however the time from pericentre to any point of the orbit is just as
important.  Both can be obtained from the action function $S_r$, whose
relationship to $S(u)$ is given below.
\begin{eqnarray}\label{eq25}
S_r  =  \int\!\! p_r dr &=& \int\! \! \dot{r} dr \nonumber\\
&=& \int \!\! \sqrt{2 \varepsilon + 2Ar^{-\alpha} -h^2 r^{-2}}\  dr \nonumber\\
&=&  h \int \!\! \sqrt{2 \varepsilon h^{-2}r^2 +2 Ah^{-2} r^k - 1}\ r ^{-1}
  dr \nonumber \\
&=&  k^{-1} h \int^{u_p}_u \!\!\sqrt{S(u)}\  u^{-2} du\;.
\end{eqnarray}
If we now use our quadratic approximation we find
\[
S_r = qk^{-1} h \int^{u_p}_u\!\!\sqrt{e^2 \bar{u}^{2}
  -(u-\bar{u})^2}\ u^{-2} du\;;
\]
setting $u= \bar{u} \left( 1+ e \,\cos \eta \right)$, this becomes,
setting $f= 1/e$,
\[S_r= qk^{-1} h \int^\eta_0 \!\!{\sin^2 \eta \over \left( f + \cos
  \eta \right)^2} \ d \eta \;.
\]
Now the related integral
\begin{eqnarray*}
\int^\eta_0 \!\!{(1 - \cos^2 \eta) \,d \eta \over f + \cos\eta }  
\eqnsp&=&\eqnsp   \int^\eta_0 \!\! \left( {1-f^2 \over f + \cos\eta} + f - \,\cos\eta\right) \,d \eta\nonumber\\  
\eqnsp&=&\eqnsp  -2 \sqrt{f^2-1} \tan^{-1} \left[ \sqrt{f-1 \over
    f+1} \tan \left ( {\eta \over 2} \right) \right ]\\
&&\hspace{1cm} + f \eta - \sin
  \eta\;,
\end{eqnarray*}
and the integral that we want is just $-d/df $ of this, so

\begin{eqnarray}
\int^\eta_0\!\! {\sin^2 \eta \over \left(f+\cos\eta \right)^2 }d
\eta \eqnsp&=&\eqnsp - \eta  +  {2f \over \sqrt{f^2 -1}} \tan^{-1}
\left[\sqrt{{f-1 \over
    f+1}} \tan {\eta \over 2} \right]\nonumber\\
&&\hspace{1.5cm} +  {\sin \eta \over f + \cos
 \eta}~,
\end{eqnarray}
so, putting this in $S_r$ and remembering that $f=1/e$,

{\setlength{\arraycolsep}{0.3pt}
\begin{eqnarray}\label{eq26}
 S_r =   qk^{-1} h 
\Big[ -\eta  + \frac{2}{\sqrt{1-e^2}}\,\tan^{-1} \left( \sqrt{1-e
       \over 1+e} \tan {\eta \over 2}\right)\nonumber\\ 
\hspace{1cm} + \frac{e\,\sin\eta}{1+e\,\cos\eta} \Big] \;.
\end{eqnarray}
The adiabatic invariant is given by
\begin{equation}\label{eq27}
J_r={1 \over 2 \pi} \oint p_r dr = \left({1 \over \sqrt{1-e^2}}-1
\right) qk^{-1} h\;.
\end{equation}
Now
\begin{eqnarray*}
\partial J _r/ \partial \varepsilon |_h = {1 \over 2 \pi} \oint {1
  \over \dot{r}} dr = P_r / (2 \pi)~,
\end{eqnarray*}
where $P_r$ is the radial period while
\begin{eqnarray*}
-\partial J_r/\partial h |_\varepsilon = {1 \over 2 \pi}
\oint hr ^{-2} {1 \over \dot{r}} dr = {1 \over 2 \pi} \oint \dot{\phi} dt =
\Phi /(2\pi)\;.
\end{eqnarray*}
$J_r/h $ is a function of eccentricity, so the partial differentiation is
best done via 
\[\partial (J_r/h) / \partial \varepsilon = d (J_r/h)
/de \;(\partial e/ \partial \varepsilon)_h
\]
 and 
\[
\left( \partial J_r / \partial h \right) _\varepsilon = J_r/h + \frac{d}{de}
\left( J_r/ h \right) h \left( { \partial e \over \partial h}
  \right) _\varepsilon
\]

\begin{equation}\label{eq28}
\left(\partial S_r / \partial \varepsilon \right)_h = \int\!\!\ {dr
  \over \dot{r}} =t ~,
\end{equation}
so this expression gives the time to any chosen point in the orbit.
In practice, $S_r/h$ is a function of $e$ and $\eta$ so the partial
derivative is done using 
\[
\left( \partial S_r / \partial \varepsilon \right)_h = \left( dS_r/de
\right)_h \;\left(\partial e/ \partial \varepsilon \right) _h~.
\]

In the general case, use of ${1 \over u} \propto r^k$ as a variable
does not lead to a prettier equation for $t$, such as the one Kepler
derived for $k=1$, but see the next section for logarithmic
potentials.

In the equatorial plane the total action is
$S_A=S_r+h_z \phi$.  The action variables are 
$J_r$ and $h_z=h$.  The angle variables are the phases of the
oscillations in $r$ and $\phi$ and are given by $
w_r= \partial S_A/\partial J_r$ and $w_\phi =\partial S_A/\partial
h$.  
For general orbits, the action variables are most often employed when
the potential is of the more general separable form;
$\psi =Ar^{-\alpha}-B\left(\theta\right)/r^2\;;~h^2$ is no longer
conserved but 
$I = h^2 - 2 B\left(\theta\right)$ is.  The general action is then
$S_A =S_r+ S_\theta + h_z\phi$ with $ S_\theta = \int^\theta_{\pi/2} \sqrt{I-
  h^2_z {\mathrm{cosec}}^2 \theta}~ d \theta$ 
and $J_\theta= {1 \over 2 \pi} \oint \partial S_\theta/\partial
  \theta~ d \theta.$
The angle variables are 
$w_\theta = \partial S_A/\partial J_\theta$ and $w_\phi =
\partial S_A/\partial h_z$.

\subsection{Logarithmic Potentials $\bmath{\alpha \rightarrow 0}$}
\label{subsec:log_pot}

For small $\alpha$ we write $\psi = Ar^{-\alpha} = Ar^{-\alpha}_0
e^{-\alpha \ln (r/r_0)}$ and expand to obtain $\psi =Ar_0^\alpha
\left[1- \alpha \ln \left ( r/r_0\right) + 0 \left( \alpha^2
  \right)\right]~.$  We set $A= V^2 /\alpha$ and consider
taking the limit as $\alpha \rightarrow 0$ while keeping $V^2$ fixed
  so $A$ tends to infinity.  To keep a finite potential, we have to
  subtract the constant $Ar_0^{-\alpha}$ from $\psi$, so we obtain a new
  potential
\[
\Psi = \psi - Ar_0^{-\alpha} = - V^2 \ln \left(r/r_0 \right)\;.
\]

To apply the methods of section \ref{subsec:gen_pot}, we define
\[
u=h^2/\left(V^2 r^2 \right)
\]
and consider orbits defined by pericentric and apocentric distances
$r_p$ and $r_a$.  In place of equations (\ref{eq15}) and (\ref{eq16})
we then have
\begin{equation}\label{eq29}
h^2 = V^2 {\ln \left(r_a/r_p \right)^2 \over r^{-2}_p - r^{-2}_a} = 
\frac{1}{2} V^2 r^2_p \left(1+e \right)e^{-1} \ln
\left({1+e \over 1-e} \right)\,\\
\end{equation}
\begin{equation}\label{eq30}
{\mathrm{so}}\;\; u_p = (1+e) \bar{u}\;, 
\;{\mathrm{where}}\;\;
\bar{u} = {1 \over 2e} \ln \left( {1+e \over 1-e} \right)\;,\;\;
{\mathrm{and}}\\
\end{equation}
\begin{eqnarray}
\varepsilon  &=& V^2{\displaystyle{r^2_a \ln \left(r_a /r_0\right)
  -r^2_p \ln \left(r_p/ r_0\right) \over r^2_a -r^2_p}}\nonumber\\
&=& {V^2  \over 4e} 
\left[ \left(1-e \right) \ln \left( 1+e\right) 
  - \left(1+e \right) \ln \left(1-e\right) 
  + 2e \ln \left({u_0 \over \bar{u}}\right)\right]\;.\nonumber\\ \label{eq31}
\end{eqnarray}
The orbital equation reads
\begin{eqnarray}
d \phi = {dr \over r \sqrt{2 \varepsilon h^{-2} r^2 - 2V^2 h^{-2} r^2
  \ln \left( r/r_0 \right) -1}} = {- du \over 2 \sqrt{ S_L(u)}}\;,\nonumber
\end{eqnarray}
where 
\begin{eqnarray}\label{eq32}
S_L(u)&=&2 Eu -u \ln \left( u_0 /u \right)-u^2\;,\nonumber\\
&=& \frac{u}{2e}\left[(1-e)\ln(1+e) - (1+e)\ln(1-e)\right]\nonumber\\ 
&&\hspace{4cm} +\, u\,\ln\left(\frac{u}{\bar{u}}\right) - u^2
\end{eqnarray}
and $E = \varepsilon/ V^2$ which is given in terms of $e$ via
(\ref{eq31}).
We now approximate $S_L(u)$ by the quadratic (\ref{eq20}), which shares
the same zeros.

\begin{figure}
\includegraphics[width=0.5\textwidth]{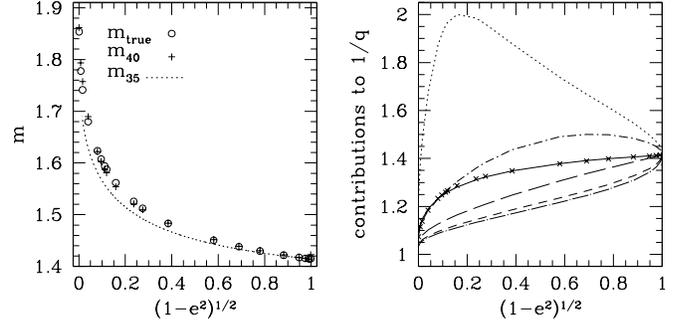}
\caption{The left panel shows a comparison between the true value of
  $m$ (open circles) and that estimated from
  equation (\ref{eq34}), which gives $m$ to better than $2\%$, 
  and the 8-point average in
  equation (\ref{eqn:8-pt}), which gives $m$ to better than $1\%$.
  The right panel shows (in thick solid) the true value of $1/q = k/m$
  overlaid (in crosses) with that derived from the 8-point average.
  The agreement is clearly good to a fraction of a percent.  The
  individual contributions made by each of the points given in the
  terms of equations (\ref{eqn:4-pt}) and (\ref{eqn:8-pt}) are also
  shown.  From the top, the various lines correspond to $u_a$
  (dotted), $u_+$ (dot short-dash), $\bar{u}$ (long dash), $u_-$
  (short dash) and $u_p$ (dot long-dash).}
\label{fig:mandS}
\end{figure}

As in section \ref{subsec:gen_orbits}, a useful analytic expression
for $q$ is given by again equating
\begin{eqnarray*}
\int_{u_a}^{u_p}S_L(u)u^{-3/2}du &=&
4\bar{u}^{1/2}\left(\sqrt{1+e}-\sqrt{1-e}\right)\times\\
&&\hspace{2cm}\left[\frac{\bar{u}}{3}(2+\sqrt{1-e^2})-1\right] 
\end{eqnarray*}
to
\begin{eqnarray*}
\int_{u_a}^{u_p}S_Q(u)u^{-3/2}du &=&
\frac{8}{3}q^2\bar{u}^{3/2}\left(\sqrt{1+e}-\sqrt{1-e}\right)\times\\
&&\hspace{3cm}\left(1-\sqrt{1-e^2}\right)\;, 
\end{eqnarray*}
where $\bar{u}$ is given by (\ref{eq30}).  This yields
\begin{equation}\label{eq34}
q^2 = \frac{2+\sqrt{1-e^2} -
  3/\bar{u}}{2\left(1-\sqrt{1-e^2}\right)}\;;
\hspace{.5cm} m=2q\;.
\end{equation}
This expression is good to $2\%$.
Once again, more accurately we calculate $q^{-1}$ from the average of
$\sqrt{S_Q/S_L}$ over $\eta$, where $u=\bar{u}(1+e\cos\eta)$.  At
$u_p$,
\begin{eqnarray}
2E = \ln(u_0/u_p)+u_p\;;\nonumber\\
dS_L/du|_p = 2E - \ln(u_0/u_p) + 1 - 2u_p = 1 - u_p\;;\nonumber\\
dS_Q/du|_p = -2q^2\bar{u}e\;;\nonumber
\end{eqnarray}
\begin{eqnarray}
{\mathrm{so}}\;
\sqrt{S_Q/S_L}\big|_p \rightarrow q\sqrt{2\bar{u}e/(u_p-1)} = q\mu_p\;,
\end{eqnarray}
\begin{equation}
{\mathrm{and\;at}}\;\eta\pm\pi/2\;,\;
\sqrt{S_Q/S_L} = q\bar{u}e{\bar{S}}^{-1/2}\;,
\end{equation}
where $\bar{S} = S_L({\bar{u}})$.  As found in section
\ref{subsec:gen_orbits}, the apocentre once again poses a problem and
we evaluate
\begin{equation}
\mu_a = \sqrt{\frac{(u_p-u)(u-u_a)}{S_L(u)}}
\end{equation}
at $u=\bar{u}(1+\lambda e)$ where $\lambda = -0.990$,
as before, 
and where $S_L(u)$ is given by (\ref{eq32}).
So a 4-point estimate of $q$ is given by 
\begin{equation} \label{eqn:4-pt}
q^{-1} \simeq q_4^{-1} = \left[\mu_p + \mu_a +2\bar{u}e\bar{S}^{-1/2}\right]/4\;,
\end{equation}
and a 8-point estimate is likewise
\begin{equation} \label{eqn:8-pt}
q^{-1} \simeq q_8^{-1} = \left[4{q_4}^{-1} + \sqrt{2}\bar{u}e(S_+^{-1/2}
  + S_-^{-1/2})\right]/8\;,
\end{equation}
where $S_\pm = S_L(u_\pm)$ and $u_\pm = \bar{u}(1\pm e/\sqrt{2})$.
Figure \ref{fig:mandS} shows the contribution of each point in
(\ref{eqn:4-pt}) and (\ref{eqn:8-pt}) along with a comparison of the
true $m$ and the value derived from the 8-point estimate via
$m_{\ref{eqn:8-pt}}=q_8k$.  Figure \ref{fig:vconst} compares an
approximate orbit to a computed one.

\begin{figure}
\includegraphics[width=0.5\textwidth]{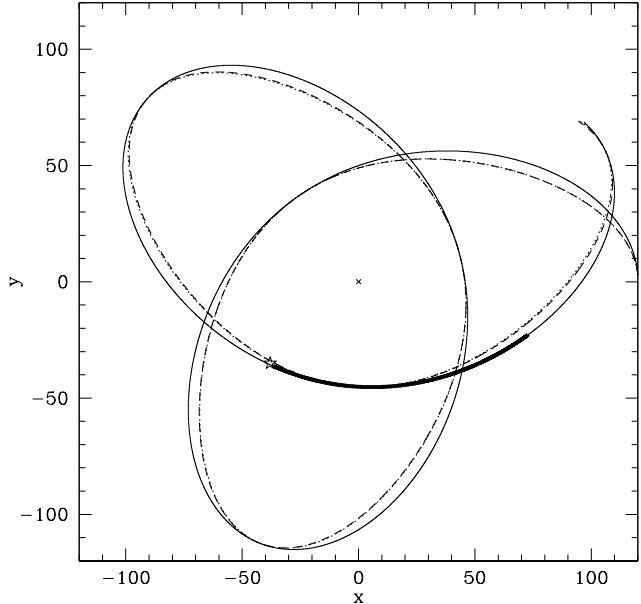}
\caption{An orbit in the logarithmic potential.  The solid line is the
  computed orbit, and the dotted line shows the approximate orbit
  $(\ell/r)^2 = 1+e\cos(m_{\rm{true}}\phi)$.  Overlaid (and hardly
  visible) is a dashed line showing an orbit with the estimated $m$,
  calculated from (\ref{eqn:8-pt}), instead of $m_{\rm{true}}$.  The
  error in this value of $m$ is less than $0.1\%$.  The heavy line
  shows the piece of such an orbit for the trailing Magellanic
  Stream, with the location of the Magellanic Clouds indicated by the
  open star.}
\label{fig:vconst}
\end{figure}

The time to a given point in the orbit is given by
\begin{eqnarray}\label{eq36}
t= \int^r_{r_p}{dr \over \dot{r}}
&\simeq& {h \over 2qV^2} \int^u_{u_p}
{-du \over u \sqrt{(\bar{u} e)^2-(u-\bar{u})^2}} \nonumber\\
&=& {h \over 2 \bar{u}
  q V^2} \int^{m \phi}_0 {d\eta \over 1+e \cos\eta} \nonumber\\
&=& {h \over
  \bar{u} q V^2 } {1 \over \sqrt{1-e^2}} \, \tan^{-1} \left[ {\sqrt{1-e
    \over 1+e}} \tan \left(q \phi \right) \right],
\end{eqnarray}
so the radial period is given by 
\begin{equation}\label{eq37}
P_r = {\pi h \over \bar{u} q V^2}\left(1-e^2\right)^{-1/2}\;.
\end{equation}

\begin{figure}
\includegraphics[width=0.5\textwidth]{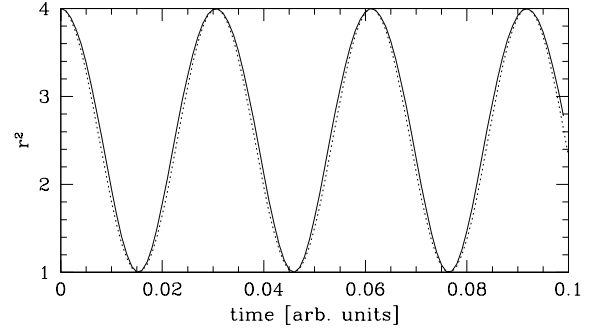}
\caption{The almost simple harmonic variation of $r^2$ as a function
  of time in an orbit with $r_a/r_p = 2$ in the logarithmic
  potential.  The dotted line is a sinusoid of the same period.}
\label{fig:harmosc}
\end{figure}

However, a much more interesting result comes from following Kepler,
whose equation comes, not from integrating the $u$ equation directly,
but by first making the substitution $v= 1/ u = V^2 h^{-2}r^2$. This gives 

\begin{eqnarray*}
t= {h \over 2q V^2 \bar{u} \sqrt{1-e^2}} \int ^v_{v_p} {dv \over
  \sqrt{\bar{v}^2 e^2 - (v- \bar{v}^2)} } = { h \over 2q V^2 \bar{u}
  \sqrt{1-e^2}} \chi\;,
\end{eqnarray*}
where we have written $v=\bar{v} (1-e\, \cos \chi)$ and
$\bar{v}^{-1} = \bar{u}(1-e^2)$; setting $\kappa = 2qV^2 \bar{u}
\sqrt{1-e^2}h^{-1}$ we see that 
\begin{equation}\label{eq38}
\chi = \kappa t\quad{\rm{and}}\quad r^2 \simeq {\displaystyle{h^2  V^{-2} \over \bar{u} (1-e^2) }}
\left[1- e\, \cos (\kappa t) \right]\;,
\end{equation}
so with this approximation $r^2$ vibrates harmonically.
Figure~\ref{fig:harmosc} shows the computed $r^2(t)$ for an orbit
together with the harmonic approximation.  The adiabatic invariant is
given by $J_r = {1 \over 2 \pi} \oint \dot{r} dr \simeq {h \over 4 \pi}
\oint \sqrt{S(u)} u^{-2} du$.  This integral was evaluated in equation
(\ref{eq26}), so for this case $k=2$ and
\begin{equation}\label{eq39}
J_r= {1 \over 2} hq \left[\left( 1-e^2 \right)^{-{1 \over 2}} -1
  \right]~.
\end{equation}

It should be emphasised that while we have set ourselves the target of
getting analytical formulae that give $m$ to 1\% or better for all
eccentricities, we have not paid attention to minimising errors in the
temporal periods.  We find that such errors are indeed higher and no
doubt our formulae could be improved upon by a study of such errors.

\section{Transformation Theory} 
\label{sec:transf_theory}

Newton (1687) realised that the ellipse was a possible orbit both in a
harmonic central potential and in an inverse square law.  In the first
case the centre of force is at the centre of the ellipse, while in the
latter case it is at the focus.  This led him to pose the question
under what circumstances can the same curve be the trajectory of a
particle under a force from one of two different centres.  Newton's
(\citeyear{1714Newton}) discussion of this is well described in
Chandrasekhar's (\citeyear{1995newt.book.....C}) book, as is later
work by \cite{1911BuAsI..28..113B}, \cite{1924Levi-Civita} and
\cite{1990Arnold}.  All demonstrate the transformation that converts
the harmonic ellipse into the Kepler ellipse and vice versa.
\cite{1981Collas} gave the relationship between equivalent
potentials. Here we show that this transformation, $S_1$, can be
embedded into a larger set of transformations that form a group.  We
mainly concentrate on the subgroup of switch transformations which
have six members but which give just three related potentials $r^{-1},
r^2$ and $ r^{-4}$ in the Kepler case, since $r^{-1}$ is self
conjugate under one of the transformations.  In the complete group,
these potentials are also related to the isochrone
\citep{1959AnAp...22..126H}.

The energy equation of a central orbit of angular momentum $h$ in a
potential $\psi (r)$ can be written

\begin{equation} \label{eq40}
{1 \over 2} \dot{r}^2 = \psi + \varepsilon - {1 \over 2} h^2 r ^{-2}~.
\end{equation}

Now consider the transformation $\tilde{r} = \rho(r)~,~ d\tilde{t}=
dt/\tau(r)$. 
Setting $F= \rho^{\prime} \tau $ we find
\begin{equation} \label{eq41}
{1 \over 2} \left({d \tilde{r} \over d \tilde{t}}\right)^2 = F^2 \left( {1
  \over 2} \dot{r}^2 \right) =F^2 \psi + F^2 \varepsilon - {F^2 \over
  2} h^2 r^{-2}~.
\end{equation}

For this to be an orbital equation like (\ref{eq40}), the 3 terms on the
right must be 
$\tilde{\psi} (\tilde{r})~, ~ \tilde{\varepsilon}$ 
and 
$ -{1 \over 2} \tilde{h}^2 \tilde{r}^{-2}$, 
but not necessarily in that
order.

The transformation 
$S_1$ 
that leads to the Newton-Levi-C\`{i}vita-Arnold
result is found by taking 
$\tilde{h}^2=h^2$, 
identifying the last
terms but switching the roles of the other two.  Thus for 
$S_1$ 
we set
\begin{eqnarray}\label{eq42}
F^2 \psi (r) = \tilde{\varepsilon}\;,
F^2 \varepsilon = \tilde{\psi}\;,
F^2  r^{-2} = \tilde {r}^{-2}\;,
\end{eqnarray}
so we obtain the transformation 
\begin{equation}\label{eq43}
\tilde{r}^2= r^2 \psi (r) / \tilde{\varepsilon}\;,
\end{equation}
with 
$\tilde{\psi}(\tilde{r})= \varepsilon r^2/\tilde{r}^2= \varepsilon
  \tilde{\varepsilon}/ \psi$.
Applying this to the power-law potential 
$\psi = Ar^{-\alpha}$ 
yields
$\tilde{r} = r^{1- \alpha/2} \sqrt{A/\tilde{\varepsilon}}$ 
so 
$\tilde{\psi}(\tilde{r})=\tilde{r}^{2 \alpha/(2-\alpha)}
\left[\varepsilon \left({\tilde \varepsilon} / A \right)^{4/(2-\alpha)}\right]$
(note: 
$\alpha=1$
gives
$\tilde{\psi}\propto \tilde{r}^2$)
the quantity in square brackets is, of course, constant.  However we
could alternatively leave the first term on the right of (\ref{eq41}),
identifying it with 
$\tilde{\psi}$
but switching the roles of the other two terms.  This leads us to the
transformation 
$S_2$
in which
\[
F^2 \psi = \tilde{\psi}~;\ F^2 \varepsilon = - {1 \over 2} \tilde{h}^2
\tilde{r}^{-2}~;\ -{1 \over 2} F^2 h^2 r^{-2} = \tilde{\varepsilon}\;,
\]
from which we deduce
$\psi r^{+2} \propto \tilde{\psi}$
and
$\tilde{r} = r^{-1} a^2$,
where
\[
a^2= {1 \over 2} {h \tilde{h} \over \sqrt{\varepsilon
    \tilde{\varepsilon}}}~.
\]
So
$S_2$
is an inversion accompanied by the change in potential
$\tilde{\psi} \propto \tilde{r}^{-2} \psi \left(a^2/\tilde{r}\right) \propto
\tilde{r}^{\alpha-2}$
(for power laws).
More generally we may ask that 
${1 \over 2} \left( {d \tilde{r} \over d\tilde{t}} \right)^2 = F^2 {1 \over 2}
  \dot{r}^2$
but that the three terms
$\tilde{\psi},~ \tilde{\varepsilon}$
and
$- {1 \over 2} \tilde{h}^2 \tilde{r}^{-2}$
that constitute this quantity are independent linear combinations of 
$F^2\psi,~ F^2 \varepsilon$
and
$F^2 \left(-{1 \over 2} h^2 r^{-2}\right)$.
If one applies two of this more general class of transformations one
after the other, it is simple to see that the net result is a
transformation of this class, that the identity transformation belongs
to the class and that every transformation has a unique inverse in the
class.  These transformations form a group in the sense of group
theory, since they clearly obey the associative law
\[
\left( T_3 T_2 \right)T_1 = T_3 \left(T_2 T_1 \right) =  T_3 T_2 T_1~.
\]

Each transformation now gives a new
$\tilde{r}(\tilde{t})$
in the new potential
$\tilde{\psi} (\tilde{r})$
 that corresponds to the old 
$r(t)$
in the old potential 
$\psi$.
To get the transformation of 
$\phi$
we remember that
$\rho^{\prime} = d \tilde{r}/dr$
and for
$S_1,~ \tilde{h} =h$
so under the transformation 
$S_1$
using (\ref{eq41}) and (\ref{eq42})
\[
d \tilde{\phi}= {\tilde{h} \over \tilde{r}^2} {d \tilde{r} \over d
  \tilde{r}/d \tilde{t}} = {h \over r} { d \ln \, \tilde{r} \over \dot{r}} =
  {d\ln \,\tilde{r} \over d\ln \,r} d \phi\;,
\]
but from (\ref{eq43})
$\tilde{r} = r \sqrt{\psi/\tilde{\varepsilon}}$
so
\[
{d\ln\, \tilde{r} \over d\ln \, r}= \left( 1 +{ 1 \over 2}\, {d\ln \, \psi
  \over d\ln \, r} \right)~.
\]
 This equation takes a particularly simple form when
$\tilde{r}$
is a power of $r$, for then $\tilde{\phi} \propto \phi$.
Furthermore, this occurs if and only if $\psi$ follows a power law in $r$;
so
\begin{eqnarray*}
\psi & = & A r^{-\alpha}\,,~d \tilde{\phi} = \left( 1 - {\alpha \over 2}
  \right) d \phi \,,~ \tilde{r} = r^{1-{\alpha \over 2}} \left( {A
    \over \tilde{\varepsilon}} \right)^{{1 \over 2}} \\
\tilde{\psi}& = & \varepsilon \left( {\tilde{\varepsilon} \over
    A}\right)r^\alpha = \varepsilon \left( {\tilde{\varepsilon}\over
    A}\right)^{2 \over k} \tilde{r}^{2 \alpha \over k}\,,~ 
\tau = \left (1-{\alpha \over 2}\right)^{-1} \left(
      {\tilde{\psi} \over \varepsilon} \right)~.
\end{eqnarray*}
Notice that the new potential depends on the energy of the old orbit,
so a pair of orbits of different energies in the old potential will
map into a pair of orbits in two {\bf{different}} new potentials that
differ by a constant factor, {\it{c.f.}} \cite{1995Rosquist}.  If we write 
$z=re^{i \phi}$
and
$\tilde{z} = \tilde{r} e^{i \tilde{\phi}}$
then, from the above, the
$S_1$
mapping is of the form
$\tilde{z}  \propto z^{1-\alpha/2}$,
i.e. a conformal map in the complex plane.  In general, a closed orbit
will map into an unclosed Lissajoux rosette, but when 
$1- \alpha/2 = N_1 /N_2$
where
$N_1$
and
$N_2$
are relatively prime integers, then the transformation of an orbit that
closes after one turn will be an orbit that closes after 
$N_2$
turns which has 
$N_1$
times as many apsides.

For $\alpha = 1$, we have the famous example that transforms Kepler's
ellipse into the simple harmonic oscillator.  This is
\[
\tilde{r}  =  r^{1 \over 2} \left( A/\tilde{\varepsilon} \right)^{1 \over 2},\quad \tilde{\phi} = {\scriptstyle{1 \over 2}} \phi ~,\quad 
\tilde{\psi}  =  \varepsilon \left( \tilde{\varepsilon}/A \right)^2
\tilde{r}^2
\]
If we apply 
$S_1$
again, this time starting with 
$\tilde{\alpha}= -2$
we find
\begin{eqnarray*}
S_1 \left[\tilde{r}\right] & \propto & \tilde{r}^2 \propto r\\
{\rm{and}}\qquad S_1 \left[\tilde{\psi}\right] & \propto &   r^{-1}
 \end{eqnarray*}
so apart from a possible rescaling, the double transformation
$S^2_1$
leads us back to the beginning.  This is true generally, not just for
power laws, since from equation (\ref{eq43}),
\[
S_1 \left[\tilde{r} \right]   \propto  \tilde{r} \sqrt{\tilde{\psi}} \propto r
\sqrt{\psi}/\sqrt{\psi}  \propto  r \quad {\rm{and}} \quad
    S_1\left[\tilde{\psi}\right] \propto \frac{1}{\tilde{\psi}} \propto
     \psi~.
\]
We shall ignore the dull rescalings in what follows and write $S_1^2 =
I$, the identity. This is in agreement with the concept that a
repeated switch leads to no transformation.

We now apply the transformation $S_1$ to orbits in one of our
potentials $\psi = A r^{-\alpha}$ with $0< \alpha < 2$.  The new
potential will be $\tilde{\psi} \propto \tilde{r}^{2 \alpha/k}$, which
will be a positive power of $\tilde{r}$ and the transformed orbit
takes the form
\[
\left({\tilde{\ell} \over \tilde{r}}\right)^2 = 1+ e \, \cos \left({2m
  \over k}\, \tilde{\phi} \right)~,
\]
which is indeed an ellipse when $m=k$ as for the Kepler case, which
transforms to the harmonic potential.  Remarkably, it is always
$\tilde{r}^{-2}$ on the left whatever $\alpha$ we start from, but the
values of $2m/k$ vary with $\alpha$.

$S_1$ is the basis for the regularization of the close encounters of
two bodies carried out in three dimensions by Kustaanheimo \& Stiefel
(\citeyear{1965Kustaanheimo}).

\subsection{The Switch Subgroup} 
If we try to find a transformation that switches the angular momentum
and potential terms in (\ref{eq41}) while leaving the energy term
unchanged, we fail because $\tilde{\varepsilon} = \mathrm{const} =
\varepsilon$ and with $F$ constant we are unable to accomplish the
desired switch.

However, we may apply first $S_1$ and then $S_2$:
\begin{eqnarray*}
S_1 \left[r \right] &=& r \sqrt{\psi \over \tilde{\varepsilon}}~~;\hspace{4mm}
\quad S_1 \left[ \psi \right] = \varepsilon \tilde{\varepsilon}/
\psi\;;\\
S_2 \left[S_1 [r] \right] &=&{\tilde{h}\, S_2 \!\left[ \tilde{h}\right]
  \over 2 \sqrt{\tilde {\varepsilon}\,S_2 \!\left[
      \tilde{\varepsilon}\right]}}\, \sqrt{\tilde{\varepsilon \over
    \psi}} {1 \over r} = 
{ \tilde{h}\,S_2 \!\left[\tilde{h}\right] \over 2
  \sqrt{S_2\!\left[\tilde{\varepsilon}\right]}}{1 \over r
  \sqrt{\psi}}\;;\\
S_2 \left[S_1 [\psi] \right] &=&\ - {2
  S_2\!\left[\tilde{\varepsilon}\right] \varepsilon \over \tilde{h}^2}\,r^2~.
\end{eqnarray*}
This transformation is not the one we obtain by applying $S_2$ first
and then $S_1$:
\begin{eqnarray*}
\tilde{r}=S_2\left[r\right]  &=&  \frac{1}{2} {h \tilde{h} \over
  \sqrt{\varepsilon \tilde{\varepsilon}}} {1 \over r}\;; 
\qquad 
S_2\left[\psi\right] = - {2 \tilde{\varepsilon} \over h^2} \psi r^2\;;\\
S_1 \left[ S_2[r] \right]  
&=&  {\tilde{h} \over \left(-2 \varepsilon S_1\left[ \varepsilon \right] \right)^{1 \over 2}}  \sqrt{\psi}~;\\
S_1\left[ S_2 [\psi] \right]  
&=&  {\tilde{\varepsilon} S_1 \left[
      \tilde{\varepsilon}\right] \over \tilde{\psi}} = -
       {\scriptstyle{1 \over 2}} h^2 S_1 \left[
         \tilde{\varepsilon}\right]/\left( \psi r^2 \right)\;.
\end{eqnarray*}
Applying this double transformation twice gives
\[
r_4 \propto 1/\left( r \sqrt{\psi} \right)\; {\mathrm{and}}\; \psi \propto r^2\;,
\]
which is the same as $S_2 S_1$ up to constants of proportionality,
while a further application of $S_1  S_2$ gives
\[
r_6 \propto r \qquad {\rm{and}} \qquad \psi_6 \propto \psi~,
\]
so the triple application of $S_1 S_2$ gives a multiple of the
identity.

Before going any further, let's see where we can get if we start with
$\psi \propto 1/r$. We can get to $\psi \propto r^2$ and back using
$S_1$, but using $S_2$ leaves $\tilde{\psi} \propto 1/\tilde{r}$, so
Newton's law is invariant under $S_2$.  However, $S_2$ acting on $r^2$
leads to $\tilde{\psi} \propto \tilde {r}^{-4}$, but a further
application of $S_1$ leaves $r^{-4}$ invariant.  Thus under the
transformations considered so far, there are conjugate orbits in the
$r^{-1},~r^{2}$ and $r^{-4}$ potentials.  More generally, if we start
with $\psi \propto r^{-\alpha}$, then $S_1$ gets us to $\psi \propto
r^{2 \alpha/(2- \alpha)}$ (where we have dropped the tildes), while
$S_2$ brings us to $\psi \propto r^{\alpha-2}$.  The double
transformations $S_2 S_1$ and $S_1 S_2$ give $\psi \propto r^{-4/(2-
  \alpha)}$ and $r^{2(2- \alpha)/\alpha}$ respectively, while $S_1S_2S_1$
leads to $\psi \propto r^{-4/\alpha}$.  Further applications only
bring us back to potentials already included.  In fact, there are six
transformations in this subgroup, yielding a conjugacy of orbits in
the six potentials $r^{-\alpha}, ~r^{2 \alpha/(2-\alpha)},~r^{\alpha
  -2},~r^{-4/(2- \alpha)}, ~r^{2(2- \alpha)/\alpha}$ and $r^{-4/\alpha}$.
For $\alpha = -1$, these are $r,~r^{-2/3},~r^{-3},~r^{-4/3},~r^{-6}$
and $r^4$.  For $\alpha = 1/2$, they are $r^{-1/2},~r^{2/3},~r^{-3/2},
~r^{-8/3},~r^6$ and $r^{-8}$.  These powers become somewhat bizarre
for small $\alpha$.
\begin{eqnarray*}
\alpha &=& 1/6\; {\rm{gives}}\;
r^{-1/6},~r^{2/11},~r^{-11/6},~r^{-24/11},~r^{22}\;
{\rm{and}}\; r^{-24}\;;\\
\alpha &=& -1/6\; {\rm{gives}}\;
r^{1/6},~r^{-2/13},~r^{-13/6},~r^{-24/13},~r^{-26}\
{\rm{and}}\; r^{24};
\end{eqnarray*}
degeneracies similar to those for the Kepler potential occur for
$\alpha =1,~ 4$ or  $\pm 2$.

The simple relationship
$\tilde{\phi}= \phi \left( 1- \alpha/2 \right)$ holds only for power
law potentials under the $S_1$ transformation.  Under $S_2$ we find
\begin{eqnarray*}
d \tilde{\phi} = \tilde{h}\tilde{r}^{2} \left( d \tilde{r}/d
\tilde{t} \right)^{-1} d \tilde{r} &=& - \left( -2 \varepsilon
\right)^{1/2} r^{-1} dr / \dot{r} \\  &=& - { \left (-2 \varepsilon
  \right)^{1/2} \over h} r d \phi\;,
\end{eqnarray*}
which no longer gives a simple relationship of $\tilde{\phi}$ to
$\phi$.  However, 
${\left ( - \tilde{\varepsilon} \right )^{1 / 4} \over \tilde{h}^{1/2}}
\int\!\! \tilde{r}^{1/2} d \tilde{\phi} = \tilde{\chi}\left(\tilde{r}
\right) = - {\left ( - \varepsilon \right)^{1/4} \over h^{1/2}} \int \!\! r^{1/2} d
\phi = - \chi $,
so if $\chi (r)$ is known for the first orbit then, with
$\tilde{r}(r)$ known, $\tilde{\chi}(\tilde{r})$ is known for the
second.

These transformations are not restricted to power-law potentials.
Under $S_1$, Plummer's potential $\psi = \mu \left( r^2 +b^2
\right)^{-1/2}$ transforms into
\begin{eqnarray*}
\tilde{\psi} = {2 \varepsilon b^2 \tilde{r}^{-2} \over 1 \pm
  {\displaystyle {\sqrt{1-{4 \mu^2 b^2 \over \tilde{\varepsilon}^2
          \tilde{r}^4}}}}}~,\;
{\mathrm{while\; under}}\; S_2\;{\mathrm{it\; becomes}}
\end{eqnarray*}
\[
\tilde{\psi} \propto {r^2 \over \left( r^2+ b^2 \right)^{1/2}} \propto
  {1 \over \tilde{r} \left( {\displaystyle{4 \varepsilon \tilde{\varepsilon}
        b^2 \over h^2 \tilde{h}^2}} \tilde{r}^2 + 1 \right)^{1/2}}~.
\]
It would be tedious to give the complete set but there are six;
$\psi,~S_1[\psi],~S_2[\psi],~S_2[S_1[\psi]],~S_1[S_2[\psi]]$ and
$S_1[S_2[S_1[\psi]]]$.

\subsection{The Larger Group}

When we ask that $F^2 \dot{r}^2 = \left( d \tilde{r}/ d \tilde{t}
\right)^2$ but in place of merely switching the terms on the right of
(\ref{eq41}) we ask that those terms are linear combinations of $\tilde{\psi}
\tilde{\varepsilon}$ and $ -{1 \over 2} \tilde{h}^2 \tilde{r}^{-2}$,
we obtain the full group of transformations.  A general transformation
of the group is then
\begin{eqnarray*}
\tilde{\psi} & = & F^2 \left( a_{11} \psi + a_{12} \varepsilon -
a_{13} {\scriptstyle{1 \over 2}} h^2 r^{-2} \right)\;,\\
\tilde{\varepsilon} & = & F^2 \left( a_{21} \psi + a_{22} \varepsilon -
a_{23} {\scriptstyle{1 \over 2}} h^2 r^{-2} \right)\;,\\
-{\scriptstyle{1 \over 2}} \tilde{h}^2 \tilde{r}^{-2} & = & F^2 \left( a_{31} \psi + a_{32} \varepsilon -
a_{33} {\scriptstyle{1 \over 2}} h^2 r^{-2} \right)\;,
\end{eqnarray*}
where $\sum_n a_{nj}=1$ for $j=1,2,3$.

If we take the particular transformation with
$a_{31} = a_{32} = a_{13} = a_{23} = a_{12} = 0$ then
$ a_{21} = 1 - a_{11}$ and $a_{22} = a_{33} = 1$
so we get, taking $\tilde{h} = h$ without loss of generality,
\begin{eqnarray*}
\tilde{\psi} = F^2 a_{11} \psi\;,
\tilde{\varepsilon} = F^2 \left[ (1- a_{11}) \psi + \varepsilon
  \right]\;,
\tilde{r}^{-2} = F^2 r^{-2}\;.
\end{eqnarray*}
The second of these gives the relationship of $\tilde{r}$ to $r$ when
the value of $F^2$ is taken from the third:
\[
\tilde{r}^2 = r^2  \left[ \left( 1-a_{11} \right) \psi + \varepsilon
  \right] / \tilde{\varepsilon}~.
\]
Taking $\psi = GM/r$ as our initial potential, we readily solve to
find 
$r ( \tilde{r} ) =  \left( {\tilde{\varepsilon} \over \varepsilon}
\right)^{1/2} \left(  \sqrt{\tilde{r}^2 + b^2} -b \right)$,
\[
{\rm{where}}\quad  b = { \left( 1- a_{11} \right) GM \over 2 \sqrt{\varepsilon
    \tilde{\varepsilon}}}~.
\]

\[
{\rm{The \  potential}}\quad \tilde{\psi} = {a_{11} GMr \over \tilde{r}^2} = a_{11}\left({\tilde{\varepsilon} \over \varepsilon}\right)^{1/2} {GM \over
  \sqrt{\tilde{r}^2 + b^2} +b}~,
\]
where we used $\tilde{r}^2 = \left( \sqrt{\tilde{r}^2 + b^2}-b \right)
\left( \sqrt{\tilde{r}^2 +b^2} + b \right)$.

The potential $\tilde{\psi}$ is the isochrone, see
\cite{1959AnAp...22..126H}.  This is the most general potential in
which all orbits can be found using only elementary functions
(trigonometric etc.) as stated by \citet*{1962ApJ...136..748E}; the
detailed proof of this was only published many years later in
\cite{1990MNRAS.244..111E}.

\section{Conclusions}
We have found crude, but useful, approximations to Abelian functions
by quadrating the expression under the surd while keeping the end
points as the constant parameters.  For our problem, these methods
give accuracies to better than 1\%.

We have shown that the $e=1$ `parabolic' orbits at the energy of
escape can be solved exactly, and we have given analytic expressions
for $m(e)$ which hold for all eccentricities $0 \leqslant e \leqslant
1$.  We have thus illuminated why Struck found these orbits to be such
good approximations at low and moderate eccentricities.

The transformation theory has allowed us to extend these results to
orbits in potentials which are positive powers of $r$ and we have
extended the transformations to form a group.

\section{Acknowledgements}
Our thanks are due to the Rijksuniversity of Groningen where this work
began while Donald Lynden-Bell was Blaauw Professor.  Contact with
Jihad Touma when this work was being prepared for publication informed
us of Struck's work and so changed the introduction substantially.
Greater clarity was infused by the referee Prof. Boccaletti.

\appendix
\section{Perturbation Theory}  
\label{sec:appendixA}

We have 
$S(u) = 2 Eu^\sigma + 2u -u^2 ~; ~ \sigma = 2(1-\alpha)/(2 - \alpha)$, 
$0 < \alpha < 2$.
We rewrite it in the forms;

\begin{eqnarray}\label{A1}
S(u) &=& q^2 \left[(\bar{u} e)^2 - (u- \bar{u})^2 \right] /
\left[1+ p(u) \right]^2 \nonumber\\
&=& q^2 \left( u_p - u \right) \left(u-u_a
\right)  / \left[1+ p(u) \right]^2~,
\end{eqnarray}
where 
$p(u)$ is the perturbation function that allows for the difference
between $S(u)$ and our quadratic approximation to it.  The orbit is
given by

\begin{eqnarray}\label{A2}
qk\; d\phi &=& q \int\!\!{-du \over \sqrt{S(u)}} = \int\!\! 
{\left(1+p(u)\right) du
  \over \sqrt{\bar{u}^2 e^2 - (u-\bar{u})^2 }}\\
&=& \int\!\! \left(1+p \left[
  \bar{u} \left( 1+ e \cos \eta \right) \right]\right) d \eta~,
\end{eqnarray}
where
$u= \bar{u} \left( 1+e \cos \eta \right)$.

Now $p$ may be expanded in a Fourier series in $\eta$:

\begin{equation}\label{A3}
p= a_0 + a_1\cos \eta + a_2\cos 2 \eta + a_3\cos 3 \eta + a_4\cos 4\eta 
+ \ldots
\end{equation}
where
$a_n={1 \over \pi} \int^{2\pi}_0 \cos \left( n \eta \right) d \eta$
for $ n \neq 0$.
We chose $q$ so that the average of $\sqrt{S_Q/S}$ over $\eta$ is 1.
This ensures that the average $<p>=0$, so $a_0=0$.

Keeping just those terms with $n\le 4$ we find that $p(\eta)$ is given
by
\begin{eqnarray}
p(0) &=& a_1 + a_2 + a_3 + a_4\\
p(\pi) &=& -a_1 + a_2 - a_3 + a_4\\
p(\pi/2) &=& -a_2 + a_4\\
p(\pi/4) &=& a_1/\sqrt{2} - a_3/\sqrt{2} - a_4\\
p(3\pi/4) &=& -a_1/\sqrt{2} + a_3/\sqrt{2} - a_4\;.
\end{eqnarray}
From these, we may deduce the following:
\begin{eqnarray}
a_1 &=& {\scriptstyle{1\over 4}}\left[p(0) - p(\pi) + \sqrt{2}\left(p(\pi/4)
    - p(3\pi/4)  \right)\right]\;;\\
a_2 &=& {\scriptstyle{1\over 4}}\left[p(0) + p(\pi) - 2 p(\pi/2)\right]\;;\\
a_3 &=& {\scriptstyle{1\over 4}}\left[p(0) - p(\pi) - \sqrt{2}\left(p(\pi/4)
    - p(3\pi/4)  \right)\right]\;;\\
a_4 &=& {\scriptstyle{1\over 4}}\left[p(0) + p(\pi) + 2 p(\pi/4)\right]\;.
\end{eqnarray}
The perturbed orbit is given by the implicit equations 
\begin{equation}
m\phi = \eta + \displaystyle\sum_{n=1}^{4}n^{-1}a_n\sin(n\eta)
\;\;{\mathrm{and}}\;\;
u = \bar{u}(1+e\cos\eta)\;.\nonumber
\end{equation}
In the above solution for the $a_n$, we have not used the combination
${\scriptstyle{1\over 2}}\left[p(\pi/4)+p(3\pi/4)\right]$ because the
condition $<p>=0$ ensures its consistency and we replace $p(\pi)$ by
$p(\phi)$ with $\phi = \cos^{-1}(-0.990)$.

\section{Unbound Orbits}

When $\varepsilon > 0$ the $E$ term dominates at large distances ($u$
small).  Indeed when $E >1$ the potential term never dominates.  We
write the equation for the orbit in the alternative forms,

\begin{eqnarray}\label{eqB1}
k\;d \phi = - {du \over \sqrt{2Eu^{\sigma} + 2u -u^2}} &=&
 -{du \over \sqrt{E^{k/\alpha} U^{2(1- \alpha)} \Sigma(U)}}\nonumber\\ 
 &=&  - {kE^{k/(2 \alpha)} dU
  \over \sqrt{ \Sigma (U)}}~,
\end{eqnarray}
\[
{\rm{where}}\qquad \Sigma \left( U \right ) = 2 + 2U^{\alpha} -
E^{k/\alpha} U^2
\]
\[
{\rm{and}} \qquad U= E^{-1 /\alpha} u^{1/k} =
E^{-1/\alpha} \left(h^2/A \right)^{1/k} r^{-1}~.
\]

When $E>1$ we use the $U$ form everywhere and approximate the
$U^\alpha$ term in $\Sigma$ as a quadratic.  We specify our orbit by the
values of the impact parameter $b = h/ \sqrt{2 \varepsilon}$ and the
value of the perihelion distance $r_p$.  The energy equation at $r_p$
is 
\[
\varepsilon = - Ar_p^{-\alpha} + {\scriptstyle{{1 \over 2}}}h^2 r_p^{-2}
\]
\[
{\rm{hence}}\qquad 1+Ar_p^{-2} \varepsilon^{-1} = b^2/r_p^2 = \beta^2\;,
\qquad {\rm{say;}}
\]
$b$ and $r_p$ are both specified and $A$ is known, so we find
$\varepsilon$ as 
\[
\varepsilon = Ar_p^{-\alpha}/\left( \beta^2 -1 \right)~,
\]
and with $\varepsilon$ now known $h$ is given by $h= b \sqrt{2
  \varepsilon}$.

From (\ref{eq6}) we can now deduce the dimensionless energy
\[
E= \left( {\varepsilon \over A} \right) \left( {h^2 \over A}
\right)^{\alpha/k} = \left[ \left( \sqrt{2} \beta \right)^\alpha /
  \left( \beta^2 -1 \right) \right]^{2/k}~.
\] 

We may also express $U$ in dimensionless combinations
\[
U= \left( E \right)^{-1/\alpha} \left[2 \beta^2 / \left ( \beta^2 -1
  \right) \right]^{1/k} \left( r_p/r \right) ~.
\]

We require our quadratic approximation to $\Sigma$ to be exact at $r_p$,
that is, $U=U_p$ and exact at $U=0$ and at the centre of the range
$U_p/2$.  Then the approximation takes the form

\begin{eqnarray}\label{eqB2}
\Sigma &\equiv&  2+2U^\alpha - E^{k/ \alpha} U^2 \simeq
\left(2 / U_p+ Q^2 U \right) \left(U_p - U\right)\\
 &=&  Q^2 \left[e^2 \bar{U}^2 - \left(U- \bar{U} \right)^2
\right]~,
\end{eqnarray}
where
\begin{eqnarray*}
Q^2 &=& \left(2/U_p \right)^2 + 2 \left(2/U_p \right)^k -
E^{k/\alpha}\;;\;\;
\bar{U} = {\scriptstyle{{1 \over 2}}}U_p - \left(Q^2 U_p
\right)^{-1}\;;\\
e^2 &=& 1+ 2 \left( Q^2 \bar{U} \right) ^{-1}\;.
\end{eqnarray*}

Thus for $E>1$ we may integrate, using this quadratic approximation to
obtain
${u \over \bar{u}} = {\ell \over r} = 1+e_{\ast} \cos m_\ast \phi$,
where
$m^2_\ast = Q^2 E^{-k/\alpha}$ and 
$\ell = E^{-1/2} \left[2 \beta^2/ \left(\beta^2-1 \right)
\right]^{1/k} \left(r_p/\bar{U} \right) = {2 \over r_p^{-1} + r_a^{-1}}$.

When $0<E<1$, the $E$ term dominates at large $r$ (small $u$) but the
$2u$ term dominates it when $1<U$.  We approximate the smaller of
these terms in each region and make sure that the two approximations
to $S(u)$ join smoothly with the same gradient at $U=1,
u=E^{k/\alpha}$.  The pericentre lies in the $U>1$ region where the
$S(u)$ form is appropriate, so we need $S(u_p)=0$.
\[
S(u) \simeq(c_0 + c_1 u) (u_p- u) = c_1 \left[e^2 \bar{u}^2 - (u-
  \bar{u})^2 \right];\;u \geqslant E^{k/\alpha},
\]
where $c_0$ and $c_1$ are constants to be determined and $e, \bar{u}$
follow them.

Since $U_p$ no longer lies in the zone where the $\Sigma$ form is used,
our former approximate form (\ref{eqB2}) for $\Sigma$
is not appropriate.  We write, instead

\[
\Sigma (U) = 2+ C_1U-C_2 U^2 = C_2 \left[e_\ast^2 \bar{U} - \left(U-
    \bar{U} \right)^2 \right]~,
\]
where $C_1$ and $C_2$ are constants.

Now looking at (\ref{eqB1}), it is $S(u)$ and $E^{k/\alpha}U^{2(1-
  \alpha)} \Sigma \left(U \right)$ that have to be continuous with a
continuous derivative at the junction point $u=E^{k/\alpha}$, where we
demand the derivatives be exact. We have $du/dU= kE^{k/\alpha}$ there.
Continuity requires:
\begin{equation}\label{eqB3}
\left( c_0+c_1 E^{k/\alpha} \right)
\left(u_p-E^{k/\alpha} \right)=E^{k/\alpha} \left(2+C_1-C_2 \right)~,
\end{equation}
whilst the condition on the derivative provides:
\begin{eqnarray}\label{eqB4}
\left( \sigma +1 - E^{k/\alpha} \right)  = c_1
\left(U_p - 2E^{k/\alpha} \right) -c_0\nonumber\\
\hspace{1cm} = k^{-1} \left[4(1-\alpha)+(3-2 \alpha) C_1 \right ] - 2C_2\;.
\end{eqnarray}

Finally, we demand that the value of the approximation be exact at
$u_p/2$.  This last condition takes a different form dependent on
whether $u_p/2$ is greater than or less than $E^{k/\alpha}$ as
different forms of approximation hold in those two regions.  Thus
\begin{eqnarray}
\label{eqB5}
2 + 2 E^{-1} \left({\scriptstyle {u_p \over 2}} \right)^{\alpha/k} -
 E^{-1} \left({\scriptstyle{u_p\over 2}} \right) ^{2/k}&&\nonumber\\  
=  2+C_1 E^{-1/ \alpha} \left ( {\scriptstyle{u_p
    \over 2}} \right )^{1/k} - C_2 E^{-2/\alpha} \left (u_p/2 \right)^{2/k}&&\nonumber\\
\;{\mathrm{for}}\; {\scriptstyle{u_p \over 2}} < E^{k/\alpha}\;,\;
{\mathrm{and}} \nonumber \\
 2^{\alpha/k}E + u_p - {\scriptstyle{1 \over 4}} u_p^2 =   \left (
  c_0 + c_1 {\scriptstyle{u_p \over 2}} \right ){\scriptstyle{u_p
    \over 2}}\;{\mathrm{for}}\;{\scriptstyle{u_p \over 2}} \geqslant
  E^{k/\alpha}~.&&
\end{eqnarray}

The four equations (\ref{eqB3}), (\ref{eqB4} i), (\ref{eqB4} ii) and
(\ref{eqB5}), in whichever form is relevant, are readily solved for
the four constants $c_0$, $c_1$, $C_1$ and $C_2$.

Our orbits can now be found in the form, $\left( {\ell / r} \right)^k
= 1+e \cos (m\phi)$ for ${h^2/Ar^k} \geqslant E^{k/2}$ where $m^2 =
k^2c_1$, but for larger $r$, we get ${\ell_{\ast} / r} = 1+e_\ast
\cos \left[m_\ast \left ( \phi + \phi _\ast \right) \right ]$.

\bibliographystyle{mn2e}

\label{lastpage}

\end{document}